\documentclass[sigconf]{acmart}

\AtBeginDocument{%
  \providecommand\BibTeX{{%
    \normalfont B\kern-0.5em{\scshape i\kern-0.25em b}\kern-0.8em\TeX}}}

\setcopyright{acmcopyright}

\copyrightyear{2020} 
\acmYear{2020} 
\setcopyright{acmcopyright}
\acmConference[MM '20]{Proceedings of the 28th ACM International Conference on Multimedia}{October 12--16, 2020}{Seattle, WA, USA}
\acmBooktitle{Proceedings of the 28th ACM International Conference on Multimedia (MM '20), October 12--16, 2020, Seattle, WA, USA}
\acmPrice{15.00}
\acmDOI{10.1145/3394171.3413729}
\acmISBN{978-1-4503-7988-5/20/10}
\usepackage{multirow}
\usepackage{amsmath}
\begin{document}
\fancyhead{}
\title{Reversible Watermarking in Deep Convolutional Neural Networks for Integrity Authentication}


\author{Xiquan Guan}
\affiliation{%
  \institution{University of Science and Technology of China}
%  \streetaddress{1 Th{\o}rv{\"a}ld Circle}
%  \country{Iceland}
}
\email{gxq13@mail.ustc.edu.cn}

\author{Huamin Feng}
\affiliation{%
  \institution{Beijing Electronic Science and Technology Institute}
}
\email{fenghm@besti.edu.cn}

\author{Weiming Zhang}
\authornote{Weiming Zhang and Nenghai Yu are the corresponding authors.}
\affiliation{%
  \institution{University of Science and Technology of China}
%  \streetaddress{1 Th{\o}rv{\"a}ld Circle}
%  \country{Iceland}
}
\email{zhangwm@ustc.edu.cn}

\author{Hang Zhou}
\affiliation{%
 \institution{University of Science and Technology of China}
% \country{India}
}
\email{zh2991@mail.ustc.edu.cn}

\author{Jie Zhang}
\affiliation{%
 \institution{University of Science and Technology of China}
% \country{India}
}
\email{zjzac@mail.ustc.edu.cn}

\author{Nenghai Yu}
\authornotemark[1]
\affiliation{%
 \institution{University of Science and Technology of China}
% \country{India}
}
\email{ynh@ustc.edu.cn}

\renewcommand{\shortauthors}{Xiquan Guan et al.}

\begin{abstract}
  Deep convolutional neural networks have made outstanding contributions in many fields such as computer vision in the past few years and many researchers published well-trained network for downloading. But recent studies have shown serious concerns about integrity due to model-reuse attacks and backdoor attacks. In order to protect these open-source networks, many algorithms have been proposed such as watermarking. However, these existing algorithms modify the contents of the network permanently and are not suitable for integrity authentication. In this paper, we propose a reversible watermarking algorithm for integrity authentication. Specifically, we present the reversible watermarking problem of deep convolutional neural networks and utilize the pruning theory of model compression technology to construct a host sequence used for embedding watermarking information by histogram shift. As shown in the experiments, the influence of embedding reversible watermarking on the classification performance is less than $\pm 0.5\%$ and the parameters of the model can be fully recovered after extracting the watermarking. At the same time, the integrity of the model can be verified by applying the reversible watermarking: if the model is modified illegally, the authentication information generated by original model will be absolutely different from the extracted watermarking information.
\end{abstract}

\begin{CCSXML}
<ccs2012>
   <concept>
       <concept_id>10002978.10002991.10002992</concept_id>
       <concept_desc>Security and privacy~Authentication</concept_desc>
       <concept_significance>500</concept_significance>
       </concept>
 </ccs2012>
\end{CCSXML}

\ccsdesc[500]{Security and privacy~Authentication}
\keywords{Reversible watermarking, Convolutional neural networks, Security, Integrity authentication}

\maketitle

\section{Introduction}

Deep convolutional neural networks (CNNs) have obtained significant achievements in computer vision recently such as image classification~\cite{he2016deep}, target tracking~\cite{li2019siamrpn} and automatic driving~\cite{casanova2018iterative}. However, the structures of the models are increasingly complex and the training of deep neural network models is difficult: several weeks are essential for a deep ResNet (ResNet152) with GPUs on ImageNet~\cite{he2016deep}. As a result, a large number of trained deep learning models have been published on the website to help people reproduce the results or improve the performance of networks by fine-tuning. During the spread of these trained models, illegal tampering has become an important issue threatening the security of the shared models. A classical method is backdoor attacks on CNNs~\cite{gu2017badnets}. The backdoor is defined as a hidden pattern injected into a deep neural network model by modifying the parameters while training. The backdoor does not affect the model’s performance on clean inputs, but forces the model to produce unexpected behavior if and only if a specific input is applied. Besides, model-reuse attacks~\cite{ji2018model} also threaten the networks. These illegal tampering will leave fatal flaws and reduce the accuracy of the trained model. Once these “infected” parent models are utilized for training, the flaws will spread like viruses in the child models and if these “infected” child models are applied in financial or security field, the flaws are likely to be exploited, which will cause destructive impact. Therefore, to ensure there is no illegal tampering on the model, that is, integrity authentication of the model, is a significant research content of model application and security.

Aimed at the security of models, there are two main protecting categories against illegal tampering: defense and authentication. Defense focuses on detection and erasure. In these methods, all models are assumed to have been tampered with illegally. Taking the backdoors defense as an example, Wang \emph{et al.}~\cite{wang2019neural} proposed \emph{Neuron Cleanse} and scanned all the model output labels to infer the potential hidden triggers. Chen \emph{et al.}~\cite{chen2018detecting} applied \emph{Activation Clustering} to detect data maliciously inserted into the training set for injecting backdoors. Liu \emph{et al.}~\cite{liu2018fine} proposed \emph{Fine-Pruning} to remove backdoor triggers by pruning redundant neurons. Most of these methods detect the backdoors passively based on the characteristics of backdoors themselves and may easily lead to missing alarm and false alarm, which will impact the performance of models after utilizing passive defense category, especially for ``clean'' models.

Another protection category is authentication which is realized by embedding some meaningful information artificially such as watermarking of CNNs. According to whether the internal details of the model are known to the public, the model watermarking can be roughly categorized into two types: white-box watermarking and black-box watermarking. \emph{White-box watermarking} embeds the watermarking information in the model internals such as weights and bias, which assumes that the internal details are public. Uchida \emph{et al.}~\cite{uchida2017embedding} propose the first CNNs watermarking technique. They choose the weights of a specific layer to embed a binary watermarking in the cover model by adding a regularization term to the loss function in the training process. Besides, Rouhani \emph{et al.}~\cite{rouhani2018deepsigns} embed the watermarking in the probability density function of the data abstraction obtained in different layers of the model. \emph{Black-box watermarking} embeds the watermarking into the model which only has application programming interface (API) access by choosing a set of key pairs to alter the decision boundary of the cover model. Yossi \emph{et al.}~\cite{adi2018turning} utilize the images with triggers and the corresponding key labels to retrain the cover model. Zhang \emph{et al.}~\cite{zhang2018protecting} propose three different generation methods of watermarking key images including choosing images in another unrelated dataset, superimposing some images from training data with additional meaning content and random noise images. Very recently, Zhang \emph{et al.}~\cite{zhang2020model} provided a watermarking framework to protect the image processing networks in black-box way.

However, these watermarking techniques are all irreversible. In the embedding process, the irreversible watermarking can only reduce the impact on the performance of the original model as much as possible, but this kind of watermarking still permanently modifies the internal parameters and destroys the integrity of the model. Therefore, this irreversible watermarking is unacceptable for integrity authentication. In order to achieve model integrity authentication, we need to propose a method which can not only embed watermarking information in the model, but also completely recover the original model parameters after extracting the watermarking, which is much more important in the models of military domain, medical domain, law application and so on. Inspired by the digital image reversible watermarking techniques, which can recover the carrier after extracting the watermarking, we proposed the first reversible model watermarking for integrity authentication of CNNs.

Generally speaking, nearly all reversible watermarking algorithms consist of two steps. First, a host sequence with a small entropy should be generated for embedding, i.e., a sharp histogram achieved by prediction errors~\cite{sachnev2009reversible}. Second, users embed the watermarking information into the host sequence by specific coding theories such as difference expansion~\cite{tian2003reversible}, histogram shift~\cite{ni2006reversible} and recursive coding~\cite{zhang2012improving}. With the development of the techniques, the coding theories have reached the optimal. So how to construct a host sequence with lower entropy is a significant research goal for reversible watermarking in images. At present, the main way of constructing host sequence is using the correlation of image pixels. Nevertheless, the characteristics of parameters in CNNs are totally different from pixels in images. Due to the incomprehensibility of the CNNs, the correlation of parameters can not be described. At the same time, the format of the parameters are different between CNNs and images. As a result, the traditional reversible watermarking methods for images can not be applied to the model directly and it is crucial to construct the host sequence which is suitable for CNNs.

To this end, we propose a CNNs watermarking method based on the pruning theory of model compression to construct the host sequence for reversible watermarking embedding. Besides, we propose a framework to realize the reversible watermarking embedding of CNNs by utilizing the coding theory learning from the images. In experiments, we take the classification networks as examples to show the effectiveness of reversible watermarking. The results of model integrity authentication is also shown in our paper. The contributions of this paper are summarized as follows:

\begin{enumerate}

\item We present a novel problem: embedding reversible watermarking into CNNs for integrity authentication.
\item We propose a method to construct the host sequence of trained model and formulate a framework to embed the reversible watermarking into CNNs by histogram shift.
\item We perform comprehensive experiments in different models to show the performance of reversible watermarking on trained models.

\end{enumerate}

\captionsetup{font={footnotesize}}
\begin{figure*}[tbp]
\begin{center}
\centering
\includegraphics[width=1\linewidth]{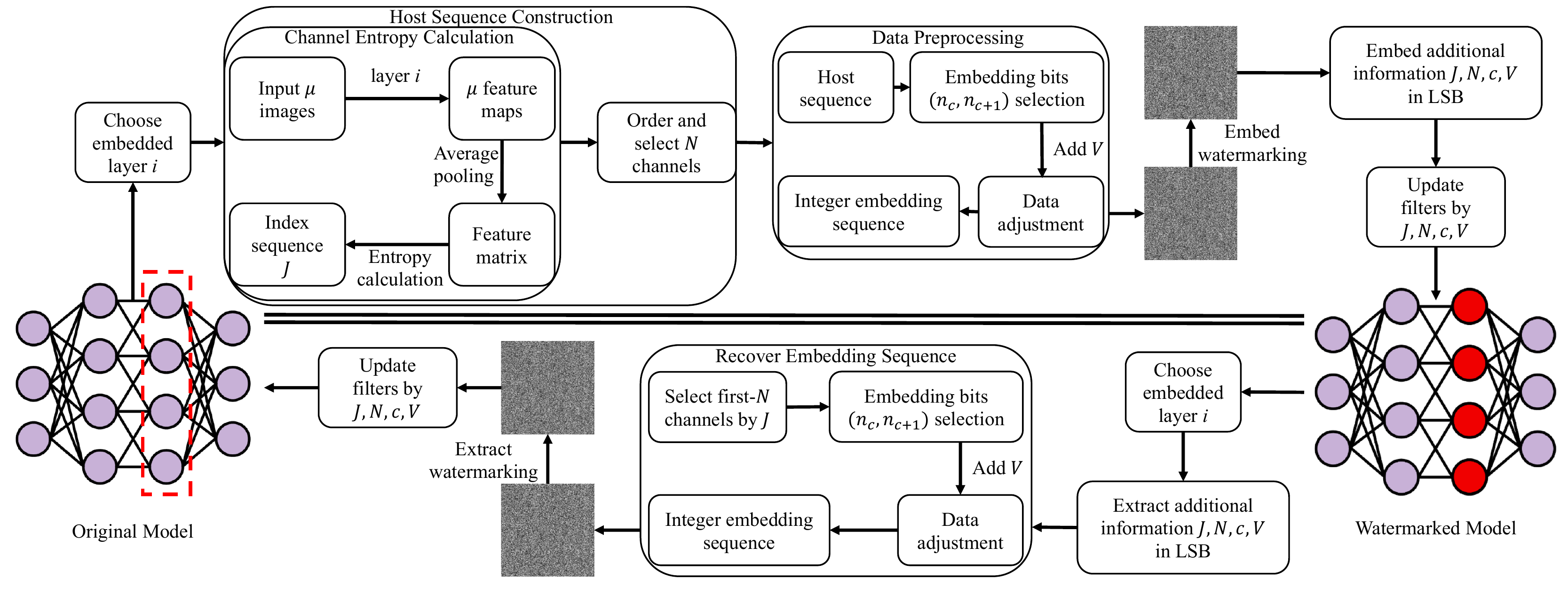}
\end{center}
\caption{\footnotesize{Reversible watermarking framework in CNNs.}}
\label{liucheng}
\end{figure*}

\section{Reversible Watermarking of CNNs}

\subsection{Problem Formulation}

For the convenience of description, we consider the $n$ convolution layers $\mathcal{C} = \left\{\mathcal{C}_{1}, \mathcal{C}_{2}, \cdots, \mathcal{C}_{n}\right\}$ of CNN model $\mathcal{M}$. We use a triplet $\mathcal{C}_{i}=\left\langle\mathcal{L}_{i}, \mathcal{W}_{i}, *\right\rangle$ to define the $i$-$th$ convolution layer, where $\mathcal{L}_{i} \in \mathbb{R}^{c \times h \times w}$ is the input tensor of layer $i$ and $\mathcal{W}_{i} \in \mathbb{R}^{d \times c \times k \times k}$ is the weights of all filters in layer $i$. The $*$ denotes the convolution operation. $c$ and $d$ denote the number of input channels and output channels respectively. $h$ and $w$ denote the height and width of input and $k$ is the size of convolution kernel.

The target of reversible watermarking embedding is to embed a $T$-bit vector $B \in\{0,1\}^{T}$, which is encrypted as a watermarking before, into $\mathcal{M}$ and obtain the marked model $\mathcal{M}^{\prime}$. So the task can be described as following:

\begin{equation}
\left\{ \begin{array}{l}
   \mathcal{M}^{\prime}=Emb\left( \mathcal{M},B \right)\\
   \left( \mathcal{M},B \right) =Ext\left( \mathcal{M}^{\prime} \right)\\
\end{array}\right.
\label{task}
\end{equation}
where $Emb\left( \cdot \right)$ and $Ext\left( \cdot \right)$ present the embedding algorithm and extraction algorithm, which are reversible for each other.

\subsection{Proposed Framework}

In this part, we briefly introduce the framework of reversible watermarking of CNNs. As shown in Fig.~\ref{liucheng}, the embedding process begins from original model (at the left of Fig.~\ref{liucheng}) and mainly includes three steps: host sequence construction, data preprocessing and watermarking embedding. The extraction process starts from watermarked model (at the right of Fig.~\ref{liucheng}) and is inverse to the embedding process. Next, we take the watermarking embedding process as an example to introduce the specific implementation of our proposed method.

\subsection{Host Sequence Construction}

As mentioned before, it is much more difficult to apply the traditional image reversible data hiding methods directly into CNN, that is to say, we must construct the host sequence for models utilizing their own characteristics. Inspired by the pruning theory in~\cite{luo2017entropy}, we adopt the entropy to rank the importance of the parameters, and select the parameters with small entropy to construct the host sequence. Notice that in irreversible watermarking method and entropy-based pruning theory, convolution layers are used as targets. Therefore, we also consider the convolution layers only and utilize the weight parameters to construct the host sequence for reversible watermarking embedding.

For the convolution layer $i$ in model $\mathcal{M}$, according to the regulations of CNN, each filter in layer $i$ corresponds to a single channel of its activation tensor $\mathcal{L}_{i+1}$, which is also the input of layer $i+1$. In entropy-based channel pruning theory~\cite{luo2017entropy}, the entropy should be calculated first to measure the importance of each channel. As a result, we first select $\mu$ images $\mathcal{I} = \left\{ I_{1}, I_{2}, \cdots, I_{\mu} \right\} $ from validation set as the model input. For image $I_{g} \in \mathcal{I}$ input $\mathcal{L}_{i} \in \mathbb{R}^{c \times h \times w}$ and the filter of this layer $\mathcal{W}_{i} \in \mathbb{R}^{d \times c \times k \times k}$, a corresponding activation tensor $\mathcal{L}_{i+1}^{g}$, which is a $d \times h^{'} \times w^{'}$ tensor, will be obtained obviously. Since the output feature map reflects the weights characteristics of this layer, we use the output of the layer $i$ as the basis for weight importance measurement. Here, we utilize global average pooling to convert the tensor into a $d$ vector as $f_j\in \mathbb{R}^{d}$. Therefore, each channel of layer $i$ will get a score of image $I_{g}$ in this way. In order to calculate the entropy, we input the whole images in $\mathcal{I}$ to calculate the channel score and obtain a matrix $\mathcal{F} \in \mathbb{R}^{\mu\times d}$ as following:

\begin{equation}
\mathcal{F}=\left( \begin{array}{c}
  f_1\\
  f_2\\
  \vdots\\
  f_\mu\\
\end{array} \right)  \triangleq \left( \begin{matrix}
  \mathcal{F}_{:,1},   \mathcal{F}_{:,2},   \cdots,    \mathcal{F}_{:,d}\\
\end{matrix} \right) 
\label{matrix}
\end{equation}
where $d$ is the channel number of output. For each channel $l \in \left\{ 1, 2, \cdots, d\right\}$, we take the distribution vector $\mathcal{F}_{:,l}$ as consideration to compute the entropy value. In order to get the frequency distribution, we first divide $\mathcal{F}_{:,l}$ into $m$ different bins and calculate the probability of each bin. Then, the entropy value can be calculated as follows:

\begin{equation}
H_{l}=-\sum_{r=1}^{m} p_{r} \log p_{r}
\label{entropy}
\end{equation}
where $p_{r}, r=\left\{1, 2, \cdots, m\right\}$ is the probability of bin $r$ and $H_{l}$ is the entropy of channel $l$. It should be noticed that there is a $\log\left( \cdot \right)$ function in the calculation formula of entropy Eq.~(\ref{entropy}), so the requirement of $p_{r} \neq 0$ must be satisfied. As a result, the compromise of the number of bins has to be considered. If we divide too much bins, some $p_{r}$ will become $0$ and the entropy will be meaningless. On the contrary, if $m$ is too small, the entropy of each channel will be not reflected enough. In our method, we utilize the iteration to obtain the largest number of $m$, which will ensure that the probability of each bin satisfies $p_{r} \neq 0$.

For the $d$ channels of layer $i$, we can obtain a corresponding entropy sequence $H = \left\{ H_{1}, H_{2}, \cdots, H_{d} \right\}$. According to the magnitude of entropy, we can sort the $H$ and obtain the ascending sequence $H=\left\{H_{j_1}, H_{j_2}, \cdots, H_{j_d}\right\}$ and obtain an index of the importance of channels as $J=\left\{j_{1}, j_{2}, \cdots, j_{d}\right\}$. Here we select an integer $N<d$ and utilize the channels corresponding to the top $N$ indexes in $J$ to construct the host sequence. As analyzed above, the smaller the entropy is, the less important the parameters are.

The filter weights $\mathcal{W}_{i} \in \mathbb{R}^{d \times c \times k \times k}$ of layer $i$ can be rewritten as $\mathcal{W}_{i}=\left\{ \mathcal{W}_{1}, \mathcal{W}_{2}, \cdots, \mathcal{W}_{d} \right\}$, where the elements of $\mathcal{W}_{i}$ belong to $\mathbb{R}^{c \times k \times k}$. We can sort the $\mathcal{W}_{i}$ by the first $N$ indexes in the index sequence $J=\left\{j_{1}, j_{2}, \cdots, j_{d}\right\}$ and obtain the sorting sequence $\mathcal{W}_{i}^{N}=\left\{ \mathcal{W}_{j_1}, \mathcal{W}_{j_2}, \cdots, \mathcal{W}_{j_d} \right\}$. For each $\mathcal{W}_{j_l} \in \mathcal{W}_{i}^{N}$, we define $\mathop{K}_\epsilon^{j_l} \in \mathbb{R}^{k \times k}$ as the kernel weights where $\epsilon \in \left\{1,2,\cdots,c\right\}$ and, therefore, $\mathcal{W}_{j_l}=\left\{ \mathop{K}_{1}^{j_l}, \mathop{K}_{2}^{j_l}, \cdots, \mathop{K}_{c}^{j_l}\right\}$. In order to construct the host sequence more similar with an image, we rearrange $\mathcal{W}_{i}^{N}$ as $W_{i}$:

\begin{equation}
W_{i}=\left( \begin{matrix}
  \mathop{K}_{1}^{j_1}&   \mathop{K}_{2}^{j_1}&   \cdots&   \mathop{K}_{c}^{j_1}\\
  \mathop{K}_{1}^{j_2}&   \mathop{K}_{2}^{j_2}&   \cdots&   \mathop{K}_{c}^{j_2}\\
  \vdots              &   \vdots              &   \ddots&   \vdots              \\
  \mathop{K}_{1}^{j_d}&   \mathop{K}_{2}^{j_d}&   \cdots&   \mathop{K}_{c}^{j_d}\\
\end{matrix} \right)_{N\times c}
\label{host1}
\end{equation}

Note that $\mathop{K}_{\epsilon}^{j_l} \in \mathbb{R}^{k \times k}$, so the $W$ can also be written as following:

\begin{equation}
W_{i}=\left( \begin{matrix}
  \omega _{1,1}&   \omega _{1,2}&   \cdots&   \omega _{1, k\times c}\\
  \omega _{2,1}&   \omega _{2,2}&   \cdots&   \omega _{2, k\times c}\\
  \vdots&   \vdots&   \ddots&   \vdots\\
  \omega _{k\times N,1}&    \omega _{k\times N,2}&    \cdots&   \omega _{k\times N,k\times c}\\
\end{matrix} \right)
\label{5}
\end{equation}
where $\omega _{\alpha ,\, \beta} \in \mathbb{R}$. And $W_{i}$ here is taken as the host sequence for watermarking embedding.

\subsection{Data Preprocessing}

As mentioned above, we obtain the host sequence $W_{i}$ utilizing the pruning theory. However, all the elements in matrix $W_{i}$ are not integer, which can not be directly applied in the traditional image reversible data hiding method. As a result, in our framework, we intercept two digits from each element of $W_{i}$ and the range of these intercepted parameters is $[-99,99]$. Then, we add $V$ to these intercepted parameters to adjust it to the appropriate range, that is, positive integer, where $V \in \mathbb{Z}$ is an adjustable parameter.

In addition to the number of interception digits, the location of interception should be considered. We assume the element $\omega _{\alpha ,\, \beta} \in W_{i}$ as following:

\begin{equation}
\omega_{\alpha ,\, \beta}=\pm \, 0 . \underbrace{00 \cdots 0}_{p \ digits} n_{1} n_{2} \cdots n_{q},
\label{host2}
\end{equation}
where $p \geqslant \text{0}$, $q>\text{0}$ and $p,q\in \mathbb{Z}$. In Eq.~(\ref{host2}), $n_{1}$ denotes the first non-zero digit of $\omega_{\alpha ,\, \beta}$, $n_{2}$ denotes the second non-zero digit of $\omega_{\alpha ,\, \beta}$ and so on. For convenience, we define the $\gamma$-$th$ non-zero digit of $\omega_{\alpha ,\, \beta}$, $n_{\gamma}$, as the $\gamma$-$th$ significant digit. It should be noticed that for different elements of $W_{i}$, the value of $p$ is different, that is, the position of the first significant digit is different.

Due to the modification of the first significant digit $n_{1}$ will cause a great influence on the value of $\omega_{\alpha ,\, \beta}$, we only consider modifying the digits from second significant digit to the last significant digit, namely, $n_{2}, n_{3}, \cdots, n_{q}$. In order to obtain a larger embedding capacity, the theory of Kalker and Willems is considered in our method. In~\cite{willems2003capacity}, the upper bound of embedding capacity under a given distortion constraint $\Delta$ was proposed as following:

\begin{equation}
\rho_{r e v}(\Delta)=\operatorname{maximize}\{E(Y)\}-E(X),
\label{upper}
\end{equation}
where $X$ and $Y$ denote the host sequence and the marked sequence after embedding respectively and $E\left( \cdot \right)$ denote entropy calculation function. According to the Eq.~(\ref{upper}), the smaller the entropy of host sequence is, the larger the embedding capacity can be obtained. Thus, we calculate the entropy of all possible host sequences constructed by intercepting \textbf{TWO} different significant digits. Then we decide the position of the selected significant digit according to the values of these entropy.

Specifically, we take the $i$-$th$ convolution layer as an example. For all the elements of $W_{i}$, $\omega_{\alpha ,\, \beta}$, we first select the second significant digit $n_{2}$ and the third significant digit $n_{3}$. After adjusting the value by $V$ mentioned before, we construct optional host sequence $W_{i}^{2,3}$. Then we count the frequency and calculate the entropy of $W_{i}^{2,3}$ as $E_{2,3}$. Similarly, we can obtain the entropy values $E_{3,4}, E_{4,5}, \cdots, E_{q-1,q}$. According to the method in~\cite{willems2003capacity}, we choose the significant digit pairs, defined as $(n_{c}, n_{c+1})$, corresponding to the minimum entropy to construct the host sequence.

Once we ensure the selection digits $(n_{c}, n_{c+1})$, we can get the integer $\omega_{\alpha ,\, \beta}^{*} = \pm n_{c} n_{c+1}$. It should be noticed that the symbols of $\omega_{\alpha ,\, \beta}^{*}$ and $\omega_{\alpha ,\, \beta}$ are consistent, that is, if the symbol of $\omega_{\alpha ,\, \beta}$ is positive, the symbol of $\omega_{\alpha ,\, \beta}$ will be positive and vice versa. Then we can obtain $\widehat{\omega}_{\alpha ,\, \beta} = \omega_{\alpha ,\, \beta}^{*} +V$ and the host sequence $\widehat{W} = (\widehat{\omega}_{\alpha ,\, \beta})_{k\times N,k\times c}$ after data processing.

\captionsetup{font={footnotesize}}
\begin{figure*}[tbp]
\begin{center}
\centering
\includegraphics[width=0.8\linewidth]{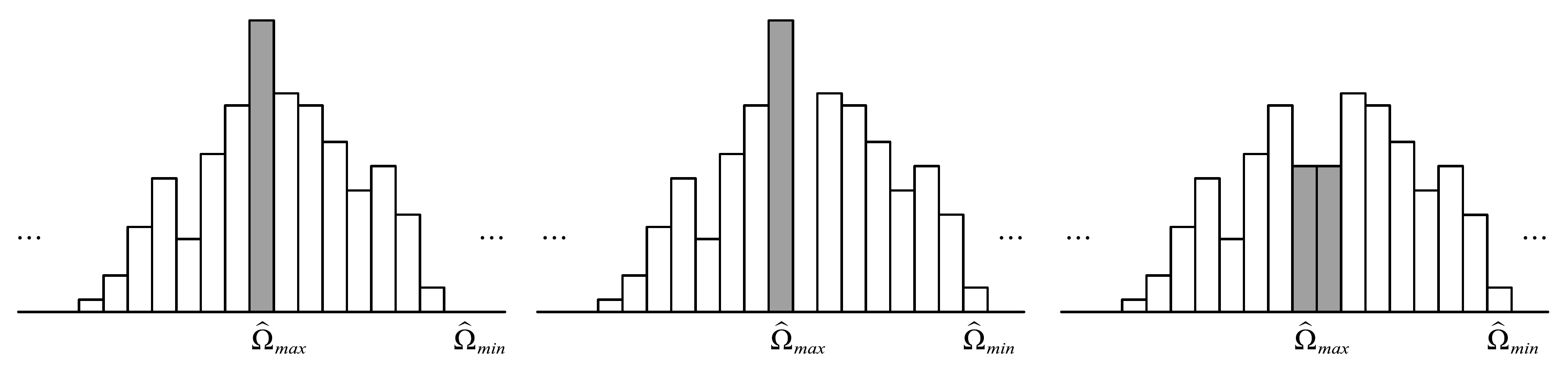}
\end{center}
\caption{\footnotesize{Illustration of Ni \emph{et al.}'s method~\cite{ni2006reversible}. Here the histogram on the left is the initial histogram, the histogram in the middle is generated by shifting the bins more than $\widehat{\Omega}_{max}$ towards right by $1$ to create a vacant bin for data embedding and the histogram on the right is the histogram embedded with watermark information utilizing HS. Without loss of generality, we assume that the number of binary $0$ and the number of binary $1$ to be embedded are equal.}}
\label{HS2}
\end{figure*}

\subsection{Embedding and Extracting Strategy}

\textbf{Embedding:} The integer host sequence $\widehat{W}$ generated above can be considered as a traditional grayscale image and we can utilize image reversible data hiding strategy to embed watermarking. In this paper, we choose histogram shift (HS) strategy~\cite{ni2006reversible}. The embedding process contains the two basic steps as following:

\textbf{(1) Histogram generation:} for $\widehat{\omega}_{i,j} \in \widehat{W}$, we generate the histogram $H(\widehat{\omega})$ the same as image: counting the number of different elements in a matrix $\widehat{W}$.

\begin{figure}[ht]
\centering
% \begin{minipage}
\includegraphics[width=3.3in]{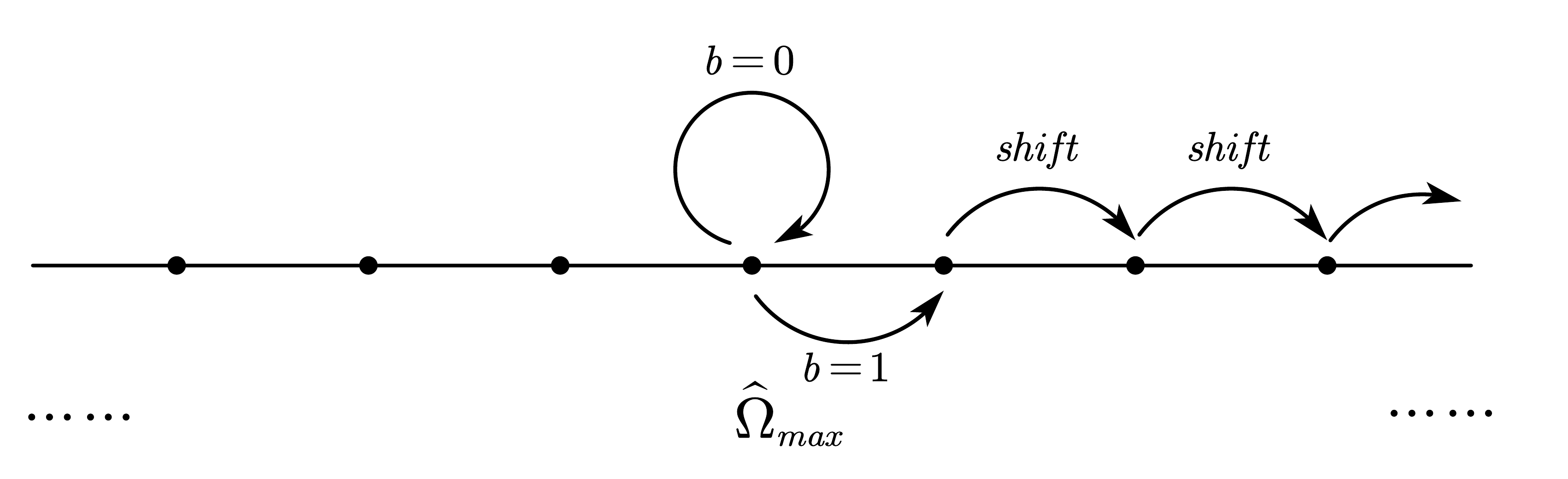}
% \end{minipage}%
\caption{Mapping rule of histogram bins described in ~\cite{ni2006reversible}: the watermark information $b$ is embedded into $\widehat{\Omega}_{max}$ and the values bigger than $\widehat{\Omega}_{max}$ in $\widehat{W}$ are shifted right while the values smaller than $\widehat{\Omega}_{max}$ in $\widehat{W}$ are remained unchanged.}
\label{HS1}
\end{figure}

\textbf{(2) Histogram modification:} we define the value in $\widehat{W}$ corresponding to the histogram peak as $\widehat{\Omega}_{max}$ and the histogram valley (generally speaking, $0$) as $\widehat{\Omega}_{min}$. Without loss of generality, in our framework, $\widehat{\Omega}_{max} < \widehat{\Omega}_{min}$. As mentioned in~\cite{Shi2016Reversible}, the HS encoding algorithm embedding one bit $b$ can be described as following:

\begin{equation}
\widehat{\omega}^{'}_{i,j}=\left\{\begin{array}{ll}
{\widehat{\omega}_{i,j}+b,} & {\widehat{\omega}_{i,j} = \widehat{\Omega}_{max}} \\
{\widehat{\omega}_{i,j}+1,} & {\widehat{\omega}_{i,j} \in (\widehat{\Omega}_{max},\widehat{\Omega}_{min})} \\
{\widehat{\omega}_{i,j}} & {\widehat{\omega}_{i,j} \notin [\widehat{\Omega}_{max},\widehat{\Omega}_{min}) }.
\end{array}\right.
\end{equation}
As shown in Fig.~\ref{HS2} and Fig.~\ref{HS1}, through embedding algorithm, watermark information is embedded into host sequence by histogram shift. 

After embedding the watermarking information, the matrix $\widehat{\omega}^{'}_{i,j} \in \widehat{W}^{'}$ is generated and we can replace the original $\mathcal{W}_{i}$ as $\mathcal{W}_{i}^{'}$ by $J$, $N$, $c$ and $V$, where $c$ is the position of $n_{c}$. First, we can obtain the new selection digits $(n_{c}^{'}, n_{c+1}^{'})$ of $\widehat{\omega}_{\alpha ,\, \beta}^{'}$ as $n_{c}^{'} n_{c+1}^{'} = \widehat{\omega}_{\alpha ,\, \beta}^{'} - V$. Therefore, the modified ${\omega}_{\alpha ,\, \beta}^{'}$ is shown as following:

\begin{equation}
{\omega}_{\alpha ,\, \beta}^{'} = \pm \, 0 . \underbrace{00 \cdots 0}_{p \ digits} n_{1} n_{2} \cdots n_{c}^{'} n_{c+1}^{'} \cdots n_{q},
\end{equation}
then the elements in Eq.~\ref{5} can be replaced as ${\omega}_{\alpha ,\, \beta}^{'}$ and get the modified $\mathcal{W}_{j_l}^{'}$. According to the parameter $N$ and index sequence $J$, we can replace $\mathcal{W}_{j_1}, \mathcal{W}_{j_2}, \cdots, \mathcal{W}_{j_d}$ in $\mathcal{W}_{i}$ by $\mathcal{W}_{j_1}^{'}, \mathcal{W}_{j_2}^{'}, \cdots, \mathcal{W}_{j_d}^{'}$ and obtain the update filter weights $\mathcal{W}_{i}^{'}$, that is, marked model $\mathcal{M}^{'}$.

It should be noted that the additional informations $J$, $N$, $c$ and $V$ should also be embedded in the the filters. Here we embed these bits into the last binary bit (converting parameters to binary numbers) of $\mathcal{W}_{i}^{'}$. Similar to the previous definition, we can define a matrix $\widetilde{W}$ by arranging the parameters of all channels in order as following:

\begin{equation}
\widetilde{W}=\left( \begin{matrix}
  \widetilde{\omega} _{1,1}&   \widetilde{\omega} _{1,2}&   \cdots&   \widetilde{\omega} _{1, k\times c}\\
  \widetilde{\omega} _{2,1}&   \widetilde{\omega} _{2,2}&   \cdots&   \widetilde{\omega} _{2, k\times c}\\
  \vdots&   \vdots&   \ddots&   \vdots\\
  \widetilde{\omega} _{k\times d,1}&    \widetilde{\omega} _{k\times d,2}&    \cdots&   \widetilde{\omega} _{k\times d,k\times c}\\
\end{matrix} \right)
\label{10}
\end{equation}
where $\widetilde{\omega} _{\alpha ,\, \beta} \in \mathbb{R}$. Then, we convert $\widetilde{\omega} _{\alpha ,\, \beta}$ to binary number as $\widetilde{\omega} _{\alpha ,\, \beta}^{B}$ and replace the last bits of $\widetilde{\omega} _{\alpha ,\, \beta}^{B}$ by encrypted additional informations $J$, $N$, $c$ and $V$. In order to keep the reversibility, we reserve a part of space in the head of watermarking information to store the original last bits information of those replaced $\widetilde{\omega} _{\alpha ,\, \beta}^{B}$ above.

\textbf{Extraction and Restoration:} We first extract the additional informations $J$, $N$, $c$ and $V$ from the filter in layer $i$. Then we construct the marked sequence $\widehat{W}^{'}$ using the methods in 2.3 and 2.4. Then, the same as embedding process, we generate the histogram $H(\widehat{\omega}^{'})$ and extract the embedded bit $b$ according to the following:

\begin{equation}
b=\left\{\begin{array}{ll}
{1,} & {\widehat{\omega}_{i,j}^{'} = \widehat{\Omega}_{max}+1} \\
{0,} & {\widehat{\omega}_{i,j}^{'} = \widehat{\Omega}_{max}},
\end{array}\right.
\end{equation}
and after extracting the embedding bits, the original element $\widehat{\omega}$ can be recovered as:

\begin{equation}
\widehat{\omega}_{i,j}=\left\{\begin{array}{ll}
{\widehat{\omega}_{i,j}^{'}-1,} & {\widehat{\omega}_{i,j}^{'} \in (\widehat{\Omega}_{max},\widehat{\Omega}_{min}]} \\
{\widehat{\omega}_{i,j}^{'}} & {\widehat{\omega}_{i,j}^{'} \notin [\widehat{\Omega}_{max},\widehat{\Omega}_{min}]}.
\end{array}\right.
\end{equation}

As mentioned above, we can recover the original $\mathcal{W}_{i}$ and update the filters in layer $i$ to obtain the original model $\mathcal{M}$.

%%%%%%%%%%%%%%%%%%%%%%%%%%%%
\captionsetup[table]{labelsep=space}
\captionsetup{justification=centering}
\begin{table*}[ht!]
\centering\caption{\\
\textbf{Comparison of Reversible Watermarking and Irreversible Watermarking}\textsc{: Qualitative comparison of two different watermarks.}}
\centering
\newcommand{\tabincell}[2]{\begin{tabular}{@{}#1@{}}#2\end{tabular}}
\centering
\begin{tabular}{c|c|c|c}
\hline
\multirow{3}{*}{ } & \multicolumn{3}{c}{Watermarking} \\
\cline{2-4}
~ & \multirow{2}{*}{Reversible} & \multicolumn{2}{c}{Irreversible} \\
\cline{3-4}
~ & ~ & Robust & Non-robust \\
\hline
Fragility & \checkmark &  & \checkmark \\
\hline
Robustness &  & \checkmark & \\
\hline
Reversibility & \checkmark &  & \\
\hline
Capacity & Medium & Small & Large \\
\hline
\multirow{2}{*}{Application} & Integrity & Intellectual & Covert \\
~ & authentication & property protection & Communication \\
\hline
\end{tabular}
\label{com}
\end{table*}
\captionsetup{justification=raggedright}
%%%%%%%%%%%%%%%%%%%%%%%%%%%%

\section{Experiments}

In this section, we firstly introduce the experimental settings (Sec.3.1) and compare the top-5 accuracy between our proposed method and irreversible watermarking technique modifying parameters directly (Sec.3.2). We then show the multi-embedding of reversible watermarking performance (Sec.3.3). Finally we show the process of integrity certification utilizing reversible watermarking (Sec.3.4).

\subsection{Settings}

For experiments, we adopt three pretrained networks AlexNet~\cite{krizhevsky2012imagenet}, VGG19~\cite{simonyan2014very}, ResNet152~\cite{he2016deep}, DenseNet121~\cite{huang2017densely} and MobileNet~\cite{sandler2018mobilenetv2} as the target models $\mathcal{M}$, and utilize the ImageNet validation images dataset consists of $50,000$ color images in 1000 classes with $50$ images per class to calculate the entropy of channels. For the process of host sequence construction, according to the relationship between depth of layer and model performance, we decide to choose the last three layers of these models to embed the reversible watermarking and choose the first $N=128$ channels in VGG19 and ResNet152, the first $N=32$ channels in DenseNet121, the first $N=48$, $N=64$, $N=96$ channels for the different layers in AlexNet and the first $N=320$, $N=960$, $N=960$ channels for the different layers in MobileNet to rearrange the weights parameters. The reason for the difference of $N$ value is that the weight tensors of different convolutions in different models are different. Besides, we choose $V=128$ as the adjustable parameter and $c=2$ as the selected significant digit position. The implementation is based on Python3.5 and MATLAB R2018a with the NVIDIA RTX 2080 Ti GPU.

%%%%%%%%%%%%%%%%%%%%%%%%%%%%%%%%
\captionsetup[table]{labelsep=space}
\captionsetup{justification=centering}
\begin{table*}[ht!]
\centering\caption{\\
\textbf{Top-5 Classification Accuracy on ImageNet}\textsc{: The comparison between our proposed method RW and LSBR embedded in the last three layers of three classical classification models: AlexNet, VGG19, ResNet152.}}
\centering
\newcommand{\tabincell}[2]{\begin{tabular}{@{}#1@{}}#2\end{tabular}}
\centering
\begin{tabular}{c|c|c|c|c|c}
\hline
\multirow{2}{*}{Network} & \multirow{2}{*}{Layer} & \multirow{2}{*}{Clean Model Accuracy (\%)} & \multicolumn{2}{c|}{Marked Model Accuracy (\%)} & \multirow{2}{*}{Length of Watermark (bits)} \\
\cline{4-5}
~ & ~ & ~ & \ \ \ LSBR\ \cite{Song2017Machine}\ \ \  & RW\ (ours) & ~\\
\hline
%AlexNet
\multirow{3}{*}{AlexNet} & \uppercase\expandafter{\romannumeral3} & \multirow{3}{*}{75.9} & 75.7 & 75.7 & 12442 \\
\cline{2-2}
\cline{4-6}
~ & \uppercase\expandafter{\romannumeral2} & ~ & \textbf{76.0} & 75.8 & 49766 \\
\cline{2-2}
\cline{4-6}
~ & \uppercase\expandafter{\romannumeral1} & ~ & \textbf{75.8} & 75.6 & 22118 \\
\hline
%VGG19
\multirow{3}{*}{VGG19} & \uppercase\expandafter{\romannumeral3} & \multirow{3}{*}{81.1} & 80.9 & \textbf{81.2} & 88474 \\
\cline{2-2}
\cline{4-6}
~ & \uppercase\expandafter{\romannumeral2} & ~ & 81.1 & 81.1& 88474 \\
\cline{2-2}
\cline{4-6}
~ & \uppercase\expandafter{\romannumeral1} & ~ & \textbf{81.0} & 80.8 & 88474 \\
\cline{1-6}
%ResNet152
\multirow{3}{*}{ResNet152} & \uppercase\expandafter{\romannumeral3} & \multirow{3}{*}{85.9} & 85.5 & \textbf{85.7} & 88474 \\
\cline{2-2}
\cline{4-6}
~ & \uppercase\expandafter{\romannumeral2} & ~ & 85.5 & \textbf{86.0} & 88474 \\
\cline{2-2}
\cline{4-6}
~ & \uppercase\expandafter{\romannumeral1} & ~ & 85.6 & \textbf{85.9} & 88474 \\
\hline
\end{tabular}
\label{all}
\end{table*}
\captionsetup{justification=raggedright}
%%%%%%%%%%%%%%%%%%%%%%%%%%%%%%%%

\subsection{Comparison with Non-reversible Methods}

First, we organize a comparison table according to the characteristics of irreversible watermarking and reversible watermarking as shown in Fig.~\ref{com}. Here we divide the irreversible watermarking into two categories, one is robust reversible watermarking, the other is non-robust reversible watermarking, similar to image steganography.

For reversible watermarking, it is fragile, reversible and the capacity is medium. It is mainly used for integrity authentication. In contrast, irreversible watermarking is irreversible. For robust irreversible watermarking, it is robust, which is utilized for intellectual property protection. For non-robust irreversible watermarking, the capacity is large. Non-robust irreversible watermarking, which is also fragile similar to reversible watermarking, is usually used for covert communication. Since we do not consider robustness and the reversible watermarking is first proposed, we only choose the two types fragile watermarking, non-robust irreversible watermarking and reversible watermarking, for comparison in the next experiment.

To illustrate the universality of our reversible watermarking method (RW), we choose a non-robust irreversible watermarking method proposed by Song \emph{et al.}~\cite{Song2017Machine}. They embedded watermarking information by least significant bit replacement (LSBR). In our experiments, we embed the watermarking in the selected layers and calculate the top-5 accuracy of classification. For convenience, we utilize \uppercase\expandafter{\romannumeral1}, \uppercase\expandafter{\romannumeral2}, \uppercase\expandafter{\romannumeral3} to represent the last layer, the second to last layer and the third to last layer. In order to make our comparative experiment more convincing, we first select the last three convolution layers of AlexNet to embed different sizes of watermark information to analyze the impact of the length of watermark information on the performance of the model. Then we choose the last three convolution layers of VGG19 and ResNet152 to embed the same size of watermark information to analyze the influence of different models on the performance of the model.

As shown in Table~\ref{all}, with same embedding bits in the same layer, the top-5 classification accuracies before or after embedding two types watermarking are almost equal ($-0.4\%\sim+0.1\%$). Besides, the accuracies between LSBR and our proposed method are almost equal ($-0.5\%\sim+0.2\%$). It should be noticed that our proposed method is reversible watermarking which can be extracted and maintain the model integrity. According to the results, embedding the reversible watermarking hardly affects the classification results of the model, which is much more different from image reversible watermarking. This can be explained in two ways. On the one hand, the modification has little influence on the value of the parameters. On the other hand, the number of parameters in these models is very large and the modification of parameters in model is limited. Besides, our method achieves the reversibility without affecting the performance of the model compared with non-robust irreversible watermarking.

\subsection{Multi-layered Reversible Watermarking}

In this part, we compare the classification performance of the models between single-layered watermarking embedding and multi-layered watermarking embedding. For the multi-layered embedding, we modify the parameters of each selected layer respectively, and then merged them into a complete modification model.

First, we choose AlexNet, VGG19 and ResNet152 to compare the effect of embedding watermark in different layers on the performance of the model. As shown in Table~\ref{muti}, the accuracies between clean model and multi-layered watermarking embedded model are almost equal ($-0.3\%\sim+0.2\%$). Then, we choose DenseNet121 and MobileNet to compare the effect of embedding watermark in single layer and multiple layers. As shown in Table~\ref{muti2}, the accuracies between clean model and embedded watermarking model are almost equal ($-0.6\%\sim-0.1\%$). As analyzed above, the embedding of single-layered watermarking has little influence on the model performance, so whether we embed multi-layered watermarking or single-layered watermarking, the performance of the models do not change much, which provides the possibility to recover the tampered model by embedding more watermarking information of model parameters' characteristic in the future.

\captionsetup[table]{labelsep=space}
\captionsetup{justification=centering}
\begin{table}[htbp]
\centering\caption{\\
\textbf{Top-5 Classification Accuracy on ImageNet}\textsc{: The results of multi-layered watermarking embedding in the last three layers of three classical classification models: AlexNet, VGG19, ResNet152.}}

\begin{tabular}{c|c|c|c}
\hline
\multirow{2}{*}{Mode} & \multicolumn{3}{c}{Classification Accuracy (\%)} \\
\cline{2-4}
~ & AlexNet & VGG19 & ResNet152 \\
\hline
Clean Model & 75.9 & 81.1 & 85.9 \\
\hline
\uppercase\expandafter{\romannumeral1}\&\uppercase\expandafter{\romannumeral2} & 75.8 & 81.3 & 85.8 \\
\hline
\uppercase\expandafter{\romannumeral1}\&\uppercase\expandafter{\romannumeral3} & 75.7 & 80.8 & 85.9 \\
\hline
\uppercase\expandafter{\romannumeral2}\&\uppercase\expandafter{\romannumeral3} & 75.8 & 81.1 & 85.9 \\
\hline
\uppercase\expandafter{\romannumeral1}\&\uppercase\expandafter{\romannumeral2}\&\uppercase\expandafter{\romannumeral3} & 75.8 & 81.1 & 85.8 \\
\hline
\end{tabular}
\label{muti}
\end{table}
\captionsetup{justification=raggedright}

\captionsetup[table]{labelsep=space}
\captionsetup{justification=centering}
\begin{table}[htbp]
\centering\caption{\\
\textbf{Top-5 Classification Accuracy on ImageNet}\textsc{: The comparison between our proposed method RW embedded in different layers and clean models }}

\begin{tabular}{c|c|c|c|c}
\hline
\multirow{2}{*}{Network} & \multirow{2}{*}{Layer} & Clean & RW & Length of Watermark \\
~ & ~ & (\%) & (\%) & (bits) \\
\hline
%Model_1
\multirow{3}{*}{DenseNet121} & \uppercase\expandafter{\romannumeral1} & \multirow{3}{*}{80.4} & 80.3 & 5530 \\
\cline{2-2}
\cline{4-5}
~ & \uppercase\expandafter{\romannumeral1}\&\uppercase\expandafter{\romannumeral2} & ~ & 80.0 & 11060 \\
\cline{2-2}
\cline{4-5}
~ & \uppercase\expandafter{\romannumeral1}\&\uppercase\expandafter{\romannumeral2}\&\uppercase\expandafter{\romannumeral3} & ~ & 80.2 & 16590 \\
\hline
%Model_2
\multirow{3}{*}{MobileNet} & \uppercase\expandafter{\romannumeral1} & \multirow{3}{*}{76.8} & 76.6 & 46080 \\
\cline{2-2}
\cline{4-5}
~ & \uppercase\expandafter{\romannumeral1}\&\uppercase\expandafter{\romannumeral2} & ~ & 76.6 & 47376 \\
\cline{2-2}
\cline{4-5}
~ & \uppercase\expandafter{\romannumeral1}\&\uppercase\expandafter{\romannumeral2}\&\uppercase\expandafter{\romannumeral3} & ~ & 76.2 & 70416 \\
\hline
\end{tabular}
\label{muti2}
\end{table}
\captionsetup{justification=raggedright}

\captionsetup[table]{labelsep=space}
\captionsetup{justification=centering}
\begin{table}[htbp]
\centering\caption{\\
\textbf{Model reconstruction error rate}\textsc{: Compare the consistency between the reconstructed model and the original model. A reconstruction rate of 0 indicates that the algorithm is completely reversible..}}

\begin{tabular}{c|c|c}
\hline
\multirow{2}{*}{Model} & \multicolumn{2}{c}{Reconstruction error rate (\%)} \\
\cline{2-3}
~ & Singe layer & Multiple layers  \\
\hline
AlexNet & 0 & 0 \\
\hline
VGG19 & 0 & 0 \\
\hline
ResNet152 & 0 & 0 \\
\hline
DenseNet121 & 0 & 0 \\
\hline
MobileNet & 0 & 0 \\
\hline
\end{tabular}
\label{muti3}
\end{table}
\captionsetup{justification=raggedright}

At the last of this subsection, we compared the difference between the original model and the reconstructed model after the extraction for the five models mentioned above. The results are shown in Table~\ref{muti3}. Both the experimental results and the theoretical analysis can prove that our method is completely reversible, that is, the integrity of the model is preserved.

%%%%%%%%%%%%%%%%
\captionsetup{font={footnotesize}}
\begin{figure*}[tbp]
\begin{center}
\centering
\includegraphics[width=0.85\linewidth]{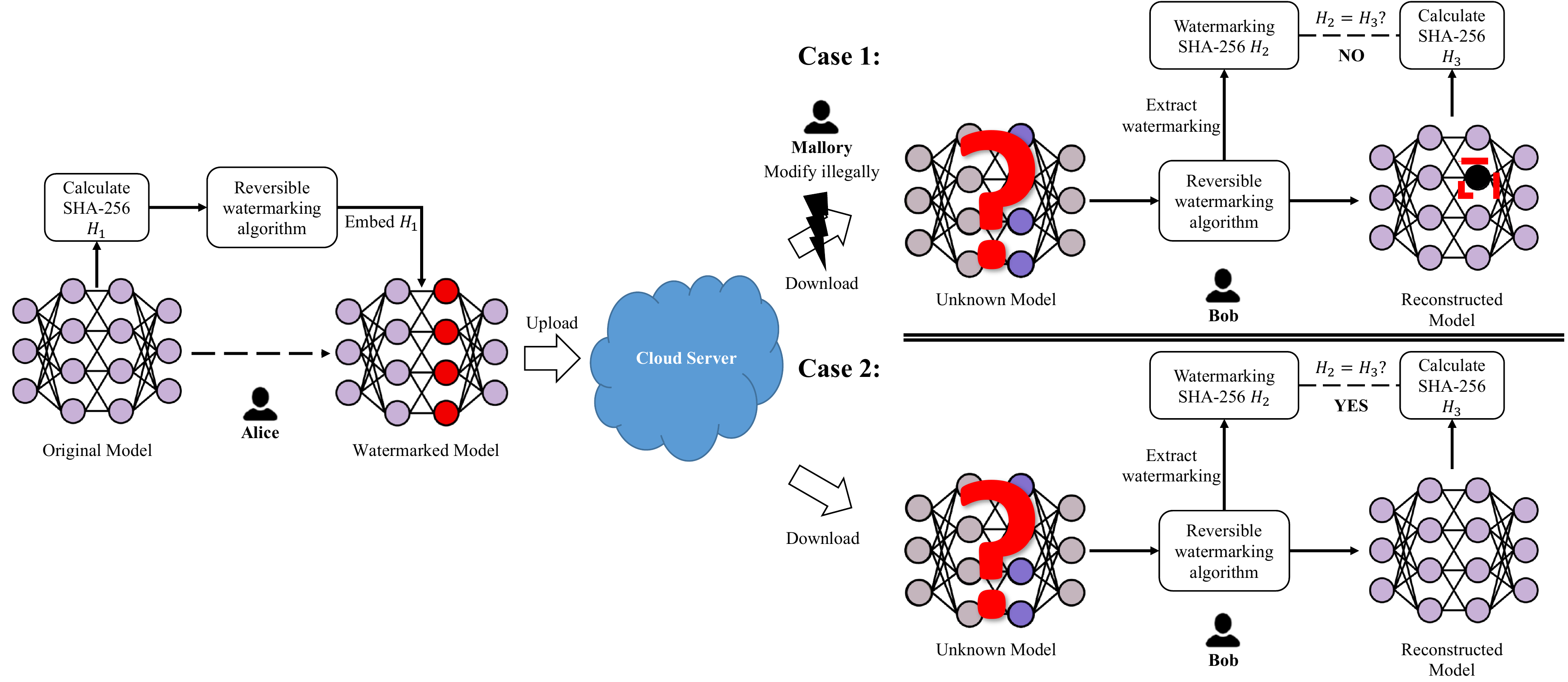}
\end{center}
\caption{\footnotesize{Integrity authentication protocol utilizing reversible watermarking of CNNs.}}
\label{renzheng}
\end{figure*}

\subsection{Integrity Authentication}

In this part, we realize the integrity authentication applied reversible watermarking. First, we utilize a Hash algorithm SHA-256 (Secure Hash Algorithm 256) to obtain the characteristic of the whole model. Then, we embed the SHA-256 value into the convolution layer by our proposed reversible watermarking algorithm. Due to the the excellent characteristics of the Hash algorithm, no matter where the attacker modifies the model, the newly generated SHA-256 value will be different from the extracted SHA-256 value.

As shown in Fig.~\ref{renzheng}, Alice is the holder of the model and she regards the SHA-256 value as the watermarking $WM_{1}$ and embed it into the model by our reversible watermarking algorithm. Then she uploads her watermarked model to cloud server for others to download. Bob downloads Alice's model from cloud server but he does not know whether the model is complete (Unknown Model), so he extracts the watermarking (defined as $WM_{2}$) from the unknown model and calculates the SHA-256 $WM_{3}$ from reconstructed model. It will be two cases here comparing $WM_{2}$ and $WM_{3}$: (1) if the model is modified illegally by Mallory, then $WM_{2} \neq WM_{3}$ (top right of Fig.~\ref{renzheng}). (2) if the model is not modified, then $WM_{2}=WM_{3}$ (bottom right of Fig.~\ref{renzheng}).

For our algorithm, we give a brief security analysis as following: We begin by presenting a definition for the security: the model is \textbf{Integrity} if it is impossible for an attacker to modify the model without being discovered. As mentioned in above, we use the SHA-256 value as the reversible watermarking to verify integrity shown in Fig.~\ref{renzheng}. Then the security of our method is reduced to the security of a cryptographic Hash algorithm (SHA-256) which is collision-resistant. Meanwhile, a security Hash function $Hash(x)$, where the domain is $X_{h}$ and the range is $Y_{h}$, is collision-resistant if it is difficult to find:

\begin{equation}
Hash(x_{1}) = Hash(x_{2}) ~~~~~~ for~~x_{1}, x_{2} \in X_{h}~~and~~x_{1} \neq x_{2}.
\end{equation}

Since the Hash function of SHA-256 is collision-resistant up to now, the method for integrity authentication is secure.

In our experiments, we choose the last convolution layer of ResNet152 to embed SHA-256 value as watermarking information. All experiments have shown that no matter where we modify or erase the parameters, our method can detect that the model has been tampered with.

\section{Conclusion}

In this paper, we present a new problem: embedding reversible watermarking into deep convolutional neural networks (CNNs) for integrity authentication. Since the state-of-art model watermarking techniques are irreversible and destroy the integrity of the model permanently, these methods are not suitable for integrity authentication. Inspired by the traditional image integrity authentication, we consider the reversible watermarking and apply it into CNNs. According to the characteristics of CNNs, we propose a method to construct the host sequence of trained model and formulate a framework to embed the reversible watermarking into CNNs by histogram shift. In the experiments, we demonstrate that our reversible watermarking in CNNs is effective and we utilize the reversible watermarking for integrity authentication in whole model.

In the future work, we will study how to determine the location where the model is modified and recover the modified parameters as much as possible by the extracted watermarking information. Furthermore, we just utilize our framework on CNNs, so we will research how to extend the reversible watermarking technique to other deep neural networks for integrity authentication.

\begin{acks}
This work was supported in part by the National Key Research and Development Program of China under Grant 2018YFB0804100, Natural Science Foundation of China under Grant U1636201 and Exploration Fund Project of University of Science and Technology of China under Grant YD3480002001.
\end{acks}

\bibliographystyle{ACM-Reference-Format}
\bibliography{sample-base}

%\appendix

\end{document}